 \pdfoutput=1
\documentclass[%
 reprint,
showpacs,
 amsmath,amssymb,
 aps,
]{revtex4-1}

\usepackage{graphicx}
\usepackage{dcolumn}
\usepackage{subfigmat}
\usepackage{epstopdf}
\usepackage{bm}

\begin{document}

\preprint{APS/123-QED}

\title[JASA-EL/Explicit wavenumber]{Explicit approximation of the wavenumber for lined ducts}
\author{Maaz Farooqui}
\author{Yves Aur\'egan}
\author{Vincent Pagneux}
\affiliation{Laboratoire d'Acoustique de l'Universit\'e du Mans, Centre National de la Recherche Scientifique (CNRS), Le Mans Universit\'e, Avenue Olivier Messiaen, 72085 Le Mans Cedex 9, France}

\date{\today} 
\date{\today}

\begin{abstract}
For acoustic waves in lined ducts, at given frequencies, the  dispersion relation leads to a transcendental equation  for the wavenumber that has to be solved by numerical methods. 
Based on Eckart explicit expression initially derived for water waves, accurate explicit approximations are proposed for the wavenumber { of} the fundamental mode {  in} lined ducts. 
While Eckart expression is 5 \% accurate, some improved approximations can reach maximum relative error of less than ${10}^{-8}$. 
The cases with  small dissipation part in the admittance of the liner and/or axisymmetric ducts are also considered.
\end{abstract}

\maketitle


\section{\label{sec:level1}Introduction}

In a duct with a locally reacting liner, a waveguide with admittance boundary conditions at the wall, the dispersion relation allows to calculate the wavenumbers as a function of the frequency 
and of the liner admittance \cite{morse1968theoretical}.
Since this dispersion relationship leads to  a transcendental equation, there is 
no closed form expression for the wavenumber and 
iterative numerical methods are most often used. 
In view of the numerous applications of lined ducts
\cite{auregan2015slow,eversman1970effect,
jones2003comparison,farooqui2016measurement,
farooqui2018,nayfeh1973acoustic,
campos2004acoustic, wenping,
rienstra2003classification,vaidya1985propagation}, it could be very interesting to have an accurate explicit
 expression of the wavenumber rather than a numerical value. 

In the field of water waves, Eckart \cite{eckart1952propagation}
\footnote{For any expression of the form, $Y = \alpha f(\alpha )$ such as $f(\alpha ) = \alpha$ for $\alpha$ close to 0 and  $f(\alpha ) = 1$ for large $\alpha$, Eckart \cite{eckart1952propagation} proposed an explicit approximation  of $\alpha$ given by $\alpha  = Y/\sqrt {f(Y)}$. For water waves, this explicit definition approximates $\alpha$ with 5\% accuracy.}
gave an approximated value of the wavenumber with 5\% accuracy on the whole frequency range.
Thereafter, other explicit approximations, extremely accurate but also more complex, were proposed for  water waves \cite{beji2013improved,simarro2013improved,vatankhah2013improved}. 
To the best of our knowledge, this type of explicit approximations of the wavenumber has not been used in acoustics
although  the dispersion relation of water waves and acoustic waves in lined ducts are very similar.
In this letter, we first show how the Eckart formula and some of its improvements can be used to compute, with an accuracy of  up to $2.8\times10^{-7}\%$, the wavenumber { of} the fundamental mode in a lined duct with a purely reactive admittance. 
Then, we present an extension of these explicit approximations from the two-dimensional (2D) case to the axisymmetric case. 
Finally, we show that we can also predict the wavenumbers when the real part of the admittance is slightly negative, modelling moderate dissipation in the liner. {  It should be noted that the proposed explicit approximations are not valid for multimodal propagation and/or highly dissipative liners}. 

%

%
\section{\label{sec:level1} Application to Non-Dissipative admittance}
\subsection{\label{sec:level1} 2D case}
\begin{figure}[ht!]
     \centering
\includegraphics[width=.75\columnwidth]{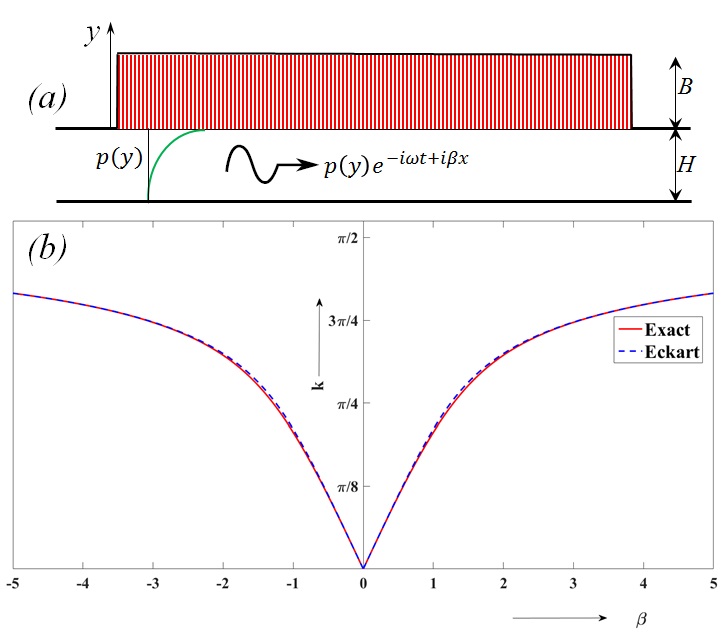}
\caption{(color online)  (a) schematic description of the 2D problem for duct of height $H$ with lining thickness $B$ . (b) Reduced frequency $k$ as a function of the wavenumber $\beta$ for lined duct. The exact solution is plotted as a continuous line while the  Eckart approximation is in dashed line ($B=H$, the liner resonance is at  $k=\pi/2$). }\label{Figure1}
\end{figure}
Let us start by considering the sound propagation in a 2D waveguide where the lower wall is rigid while the upper wall is compliant and described by a admittance $Y$ (Fig.~\ref{Figure1}(a)). When the distances are non-dimensioned by the height of the channel $H$, the Helmholtz equation, governing the propagation of the acoustic pressure $p$, is $
\Delta p + {k^2} p = 0$, where  $k =\omega H /c_0$  is the reduced frequency, $\omega$ is the frequency and $c_0$ is the sound velocity. The boundary conditions are $\partial_yp=0$ at $y=0$, for the rigid wall 
 and $\partial_y p=Y p$ at $y=1$, for the wall with liner. 
 For a uniform admittance $Y$, the modal solution can be written under the form $p = A \cosh(\alpha y) \exp(\mathrm{i}(-\omega t + \beta x))$, where 
\begin{equation}
\label{eq:0}
 \beta^2 = \alpha^2 + k^2,
\end{equation}
 leading to the dispersion relation:
\begin{equation}
\label{eq:1}
Y = {\alpha}\: \text{tanh}({\alpha}).
\end{equation}
To solve this equation with $\alpha$ as the unknown is the central subject of this paper. 
Of course, once $\alpha$ is found it yields the wavenumber $\beta$ through Eq. \eqref{eq:0}. When $Y>0$, a nice explicit approximation (coming from Eq. \eqref{eq:1})  of $\alpha$ as a function of the admittance $Y$ is given by the Eckart formula \cite{eckart1952propagation}
\begin{equation}
 \alpha_E  = \frac {Y} {\sqrt {\tanh (Y)} }.
\label{eq.5}
\end{equation}
This relationship is valid for the fundamental mode ($\alpha$ real) and it takes into account the two limit cases: $\alpha \ll 1$ for which $Y\simeq \alpha^2$ and  $\alpha \to +\infty$ for which $Y\simeq \alpha$,
giving a good approximation between these two limits.
 Eq. \eqref{eq.5} holds  for acoustic wave propagation in a 2D lined  duct up to the resonance frequency 
 of the liner (where $Y \to +\infty$). 
 This is illustrated in Fig.~\ref{Figure1}(b) where $\alpha$ and then $\beta$ are computed exactly {  as well as} by the Eckart approximation. 
 In this case,  the admittance is given by $Y= k\tan(k B)$ where $B$ is the height of the liner made of lossless tubes non-dimensioned by $H$.
 
\begin{figure}[ht!]
 \begin{subfigmatrix}{2}
  \subfigure[Eckart formula and 1-Step approximation ]{\includegraphics{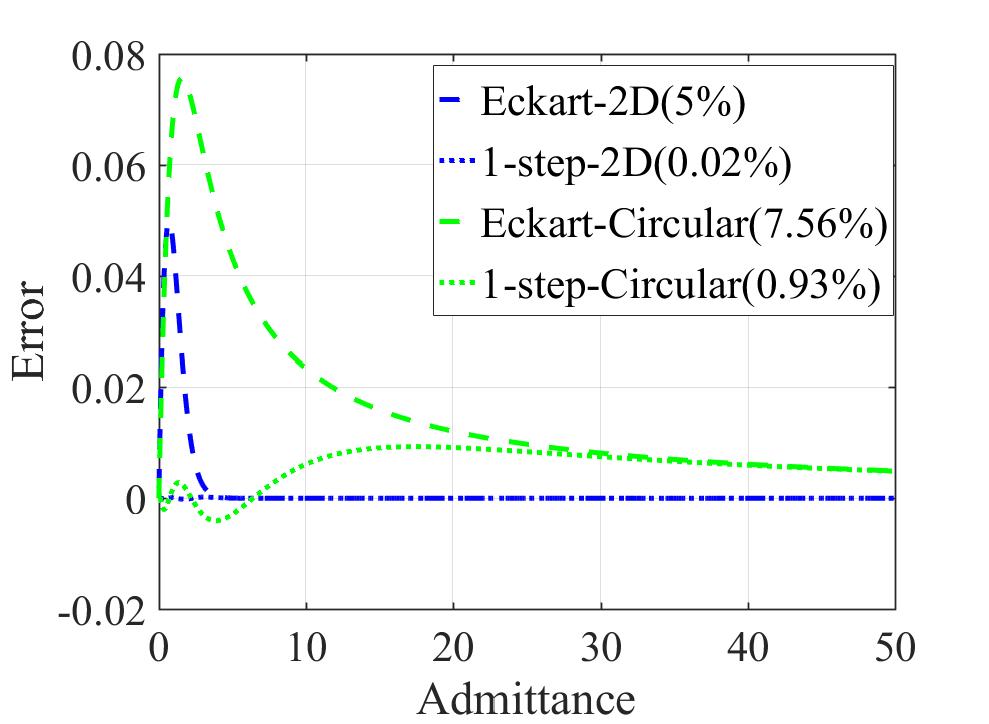}}
  \subfigure[First Newton iteration of  1-Step approximation]{\includegraphics{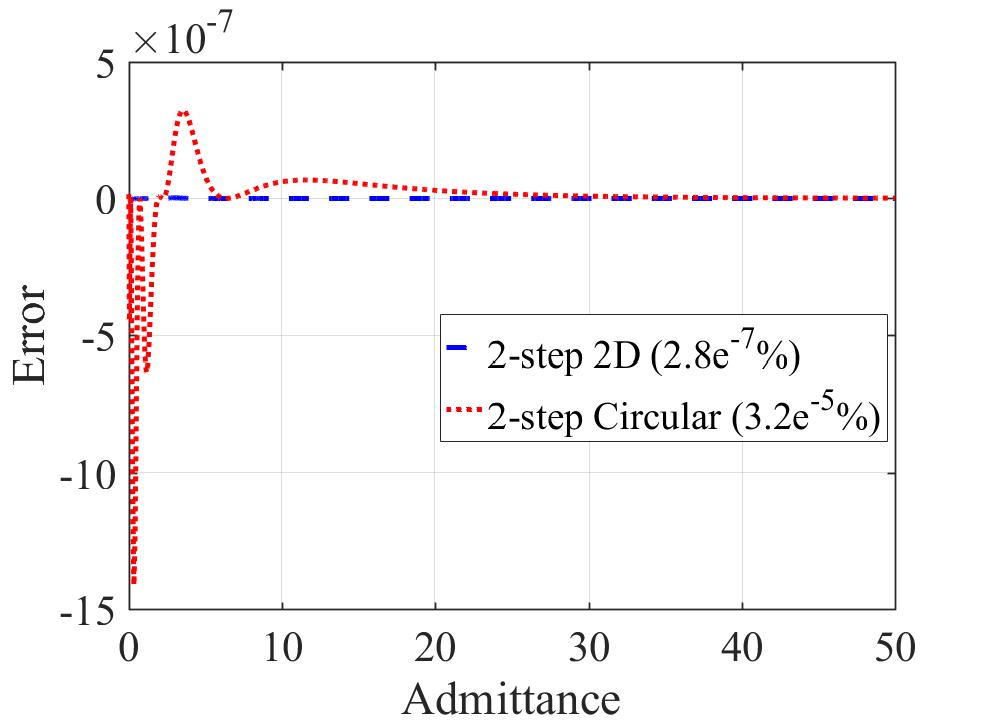}}
  \end{subfigmatrix}
 \caption{(color online)  Error  $\varepsilon$ as a function of the real admittance $Y$.  }
 \label{Figure2}
\end{figure}

 In the following, the error $\varepsilon$ on an approximate transverse wavenumber ${\alpha _ {app} }$ 
  is defined using $\varepsilon = |({\alpha _ {app} }-\hat{\alpha})/\hat{\alpha}|$ where $\hat{\alpha}$ 
  is the exact  value 
\footnote{To evaluate this error, 
we start from a given $\hat{\alpha}$. 
From the dispersion relation Eq. \ref{eq:1}, we obtain the associated admittance $Y$. 
Introducing this admittance in the approximated relation Eq. \ref{eq.5}, we obtain the approximated 
 $ \alpha_E$ and we can compute the error.}.
The error of the Eckart approximation is displayed  in blue dashed line in Fig.~\ref{Figure2}(a) as a function of the admittance $Y$. It can be seen that the error $\varepsilon$ does not exceed 5\%.

Another and much better approximation can be obtained by an empirical fit of the error of the Eckart approximation \cite{vatankhah2013improved} which leads to explicit expression
\begin{equation}
{\alpha _0} = \frac{{{{Y + {Y}^{1.986}{e^{ - (1.863 + 1.198{Y}^{1.366})}}}}}}{{\sqrt {\tanh ({Y})} }}
\label{eq.7}
\end{equation}
that is called in the following the 1-Step approximation. Using this relation, the error is reduced to $\varepsilon = 0.02\%$ and is shown in blue dotted line on Fig.~\ref{Figure2}(a) that is indistinguishable  from zero.{  The four coefficients of Eq. (\ref{eq.7}) were used as initial guess band ($\pm 1$)  in order to obtain formulas of the cases discussed later}.   

To increase further the accuracy of the prediction, it also possible to use the first iteration of the Newton method that can be written explicitly with a particularly  simple expression.  Considering $\alpha _0$  as an initial value, we obtain the 2-Step approximation\cite{simarro2013improved} :
\begin{equation}
{\alpha _*} = \frac{{{\alpha _0}^2 + {Y}{{\cosh }^2}{\alpha _0}}}{{{\alpha _0} + \sinh {\alpha _0}\cosh {\alpha _0}}}.
\label{eq1}
\end{equation}
The error in $\alpha_*$ is around $2.8\times10^{-7}\%$ and is displayed in  Fig  \ref{Figure2}(b) in dashed line.  If one intends to apply directly, the Newton's method (Eq. \eqref{eq1}) with Eckart approximation (${\alpha_E}$) instead  of $\alpha_0$, the error will be around $ 8.4\times10^{-3}\%$ 

\subsection{\label{sec:level1}Axisymmetric circular case}

Often, practical situations involve circular ducts and
the same type of work as in the 2D case can be done for axisymmetric circular ducts with liner.   
In dimensionless form, the equation governing acoustic pressure $p$ in 
axisymmetric circular ducts in the transverse direction is 
$p'' + \frac{1}{r}p' + ({k ^2} - {\beta ^2})\,p = 0$, where  $k =\omega R/c_0$  is the reduced frequency and $R$ is 
the duct radius. 
The admittance boundary condition is $\frac{{\partial p}}{{\partial r}} = Yp$ for $r=1$. For a uniform admittance, the solution is then searched under the form $p =  A {I_0(\alpha r)} \exp(\mathrm{i}(-\omega t + \beta x ))$ where $\alpha^2 = \beta^2-k^2$, leading to the dispersion relation:
\begin{equation}
Y = {\alpha}\frac{{{I_1}({\alpha} )}}{{{I_0}({\alpha })}}
\label{eq.9}
\end{equation}
where $I_{0,1}$ are the modified Bessel function of order 0 and 1. 
Following the idea of Eckart we obtain the new approximation
\begin{equation}
{\alpha_E} = \frac{Y}{\sqrt {I_1 (Y)/I_0 (Y )}} .
\label{eq1.0}
\end{equation}
The maximal error for $Y>0$ is then 7.56\%.  (see Fig.~\ref{Figure2}(a)).

 As in 2D case, one can achieve a better accuracy 
by a 1-Step approximation $\alpha_0$
\begin{equation}
{\alpha _0} = \frac{{{{Y + {Y}^{2.5285}{e^{ - (.0216 + 2.6034{Y}^{0.4615})}}}}}}{\sqrt {I_1 (Y)/I_0 (Y )}}
\label{eq1.1}
\end{equation}
with an error that is around 0.93$\%$. The corresponding 2-Step approximation $\alpha _*$ 
has a $3.2\times10^{-5}\%$ error as in Fig.~\ref{Figure2}(b):
\begin{equation}
{\alpha _*} = {\alpha _0 } + \frac{{{J_0}({\alpha _0 })(Y\,{I_0}({\alpha _0 }) - {\alpha _0 }{I_1}({\alpha _0 }))}}{{{\alpha _0 }({I_0}{{({\alpha _0 })}^2} - {I_1}{{({\alpha _0 })}^2})}}
\label{eq101}
\end{equation}
  If one intends to apply directly the Newton's method (Eq. \eqref{eq101}) with 
Eckart approximation (${\alpha_E}$) instead  of $\alpha_0$, the error will be around $ 0.01\%$.

\section{\label{sec:level1} Application in dissipative Cases}
\subsection{\label{sec:level1}2D case}
\begin{figure}[ht!]
     \centering
\includegraphics[width=.8\columnwidth]{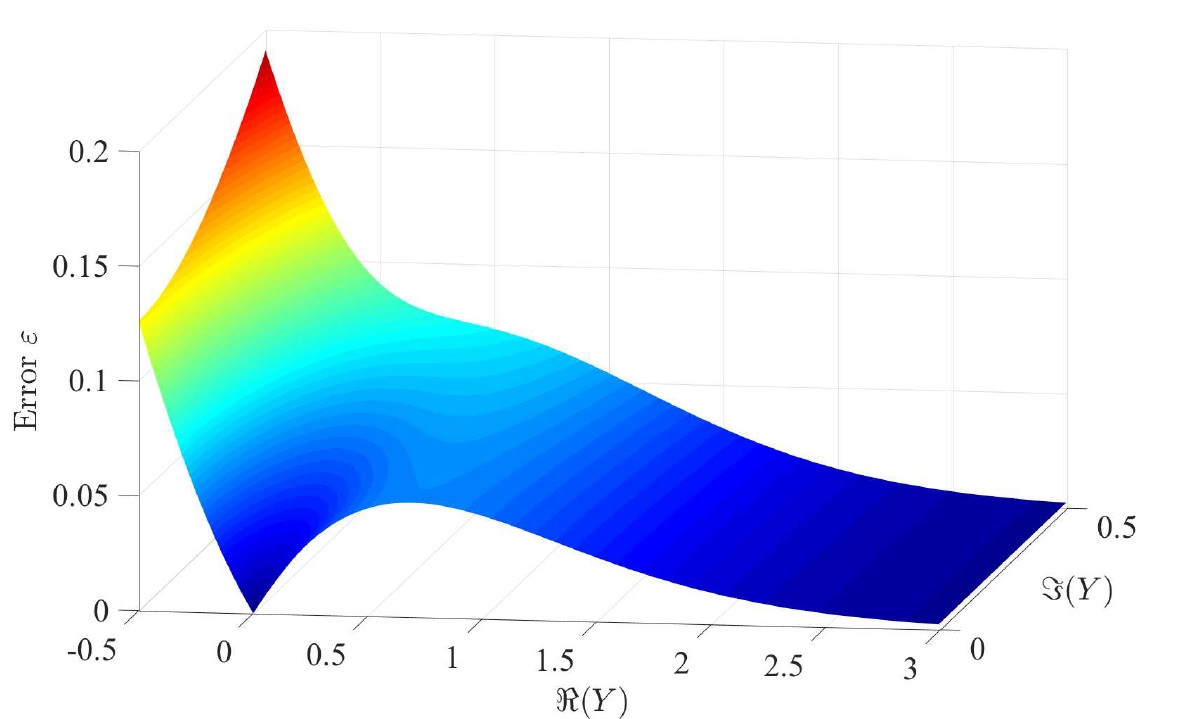}
\caption{ (color online) Error of the Eckart approximation (Eq. (3)) in the complex plane of the admittance $Y$ with
small dissipation.}
\label{Figure3}
\end{figure}

  The Eckart approximation, which is valid  for $Y$ real and positive, 
can be extended to cases of great interest for acoustics: the cases 
when the real part of the admittance is slightly negative corresponding to 
a moderate dissipation (given by a positive imaginary part of $Y$). 
The results of this continuation of the Eckart approximation in the complex plane is displayed in Fig.~\ref{Figure3} for the 2D case. 
It can be seen that this approximation is accurate around $Y=0$ and for $\Re(Y)$ large. 
When $\Re(Y)$ is negative the error increases quite rapidly with $|\Re(Y)|$. 
The effect of the dissipation is weak when $\Re(Y)$ is positive and large but adding dissipation significantly increases the error when $\Re(Y)$ is low or negative. 
Then, the worst error is 20\% corresponds to the largest negative $\Re(Y)$ and the largest $\Im(Y)$ that we have considered (Fig.~\ref{Figure4}(a)). 
Following a  procedure similar to the non-dissipative case, an improved 1-Step approximation is found as
\begin{equation}
{\hat{\alpha} _0} = \frac{{{{Y + {Y}^{2.0987}{e^{ - ( 1.8340 + 0.6507{Y}^{1.1891})}}}}}}{{\sqrt {\tanh ({Y})} }}
\label{eq143}
\end{equation}
with a maximal error  of 7$\%$  (Fig.~\ref{Figure4}(c)). 
The corresponding 2-Step approximation
$\alpha _*$ (Eq. \eqref{eq1}) has only 0.28$\%$ error (Fig.~\ref{Figure4}(e))
 for $\Re(Y) \in [-0.5,\infty ]$, $\Im(Y) \in [0,0.5 ]$.
Applying (Eq. \eqref{eq1}) with Eckart approximation ${\alpha_E}$ (Eq. \eqref{eq.5}) instead  of $\hat{a_0}$, 
the error is around $ 3.1\%$.   

\subsection{\label{sec:level1}2D Circular case}
\begin{figure}[ht!]
 \begin{subfigmatrix}{2}
  \subfigure[ $\alpha_E$(2D)($\varepsilon_{max}=19.08\%$)]{\includegraphics{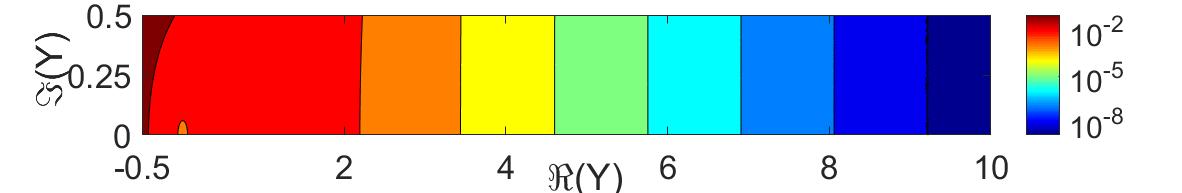}}
  \subfigure[$\alpha_E$(Circular)($\varepsilon_{max}=11.5\%$)]{\includegraphics{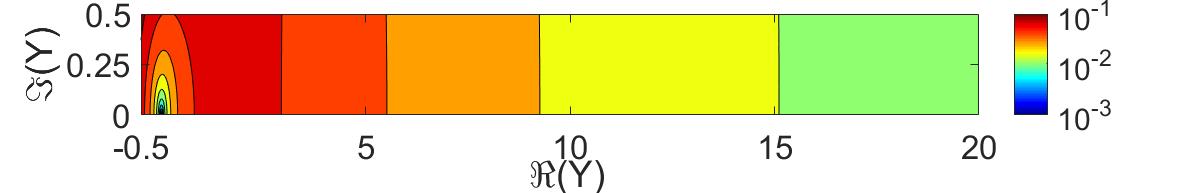}}
  \end{subfigmatrix}
   \begin{subfigmatrix}{2}
  \subfigure[Eckart formula and 1-Step approximation ]{\includegraphics{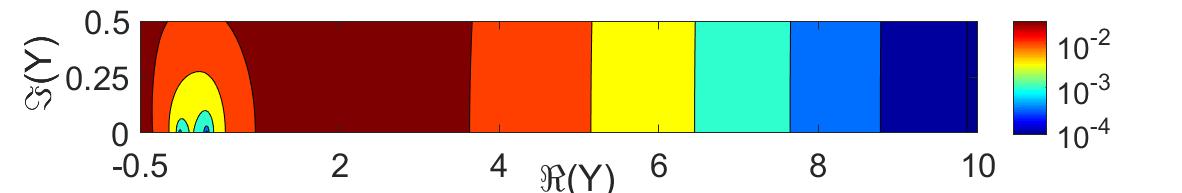}}
  \subfigure[First Newton iteration of  1-Step approximation]{\includegraphics{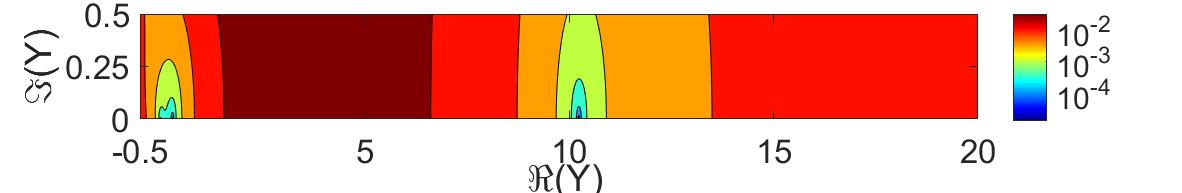}}
  \end{subfigmatrix}
   \begin{subfigmatrix}{2}
  \subfigure[Eckart formula and 1-Step approximation ]{\includegraphics{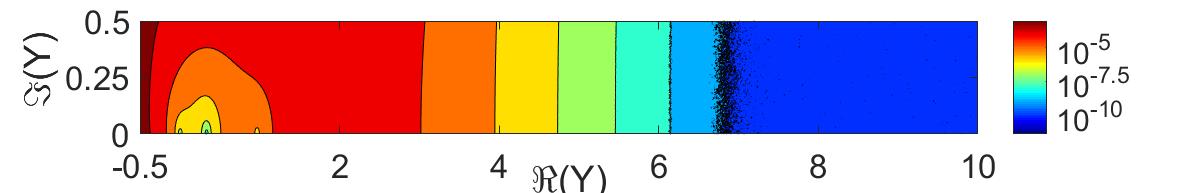}}
  \subfigure[First Newton iteration of  1-Step approximation]{\includegraphics{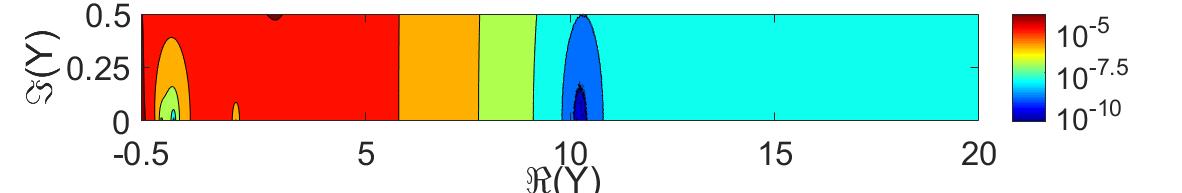}}
  \end{subfigmatrix}
 \caption{(color online)  Errors with respect to exact solution of Complex Admittance $Y$ using $Y= \alpha \tanh \alpha$ for 2D ducts and $Y = {\alpha}\frac{{{I_1}({\alpha} )}}{{{I_0}({\alpha })}}$ for Circular ducts (a,b) $\alpha_E$ from Eckart formula (c,d) $\hat{\alpha_0}$ from 1-Step proposition (e,f) $\alpha_*$ from first Newton iteration of  1-Step proposition }
 \label{Figure4}
\end{figure}

Following the same lines as previously, the Eckart approximation (Eq. \eqref{eq1.0}) error is around 11.5$\%$
 as shown in Fig.~\ref{Figure4}(b). 
 Then,   Fig.~\ref{Figure4}(d) shows that, for $Y_{Real} \in [-0.5,\infty ]$, $Y_{Imag} \in [0,.5 ]$,  
 one can achieve  6.33$\%$ error using the approximation 
\begin{equation}
{\hat{\alpha} _0} = \frac{{{{Y + {Y}^{2.0583}{e^{ - (2.0334 + 0.3717{Y}^{1.0407})}}}}}}{{\sqrt {{{{I_1}({Y} )}}/{{{I_0}({Y})}}} }}.
\label{eq198}
\end{equation}
   The error in $\hat{\alpha} _*$ (Eq. \eqref{eq101}) will be 0.02$\%$ as in Fig.~\ref{Figure4}(f). On directly applying Newton method (Eq. \eqref{eq101}) with Eckart approximation ${\alpha_E}$ (Eq. \eqref{eq1.0}) instead  of $\hat{\alpha}_0$, the error will be around $ 1\%$.

 \section{\label{sec:level1}Conclusion}
Explicit approximations for wavenumbers have been proposed to analyze the propagation in a lined duct.
The very simple Eckart expression approximates with a reasonable accuracy the exact dispersion relation  when the dissipation is weak and the imaginary part of the admittance is not too negative. It has been shown that the accuracy can be improved by introducing
 empirical corrections and/or by using the explicit first iteration of the Newton method.
These explicit approximations may be used easily for practical purposes:
explicit one-mode determination can simplify models based on low frequency acoustic wave propagation and several applications, in 2D ducts   \cite{auregan2015slow,eversman1970effect,
jones2003comparison,farooqui2016measurement,
farooqui2018,nayfeh1973acoustic} as well as Axisymmetric ducts \cite{campos2004acoustic,wenping,rienstra2003classification,vaidya1985propagation,nayfeh1973acoustic}, can be made simpler using this type of approximation at least as a starting point.

\begin{acknowledgments}
   This work was supported by the International ANR project FlowMatAc a co-operation project between France and Hong Kong. (ANR-15-CE22-0016-01)
\end{acknowledgments}


\end{document}